\documentclass[a4paper,12pt]{article}

\date {}
\usepackage{amssymb,amsfonts,amsmath,epsfig,float,graphicx,subfigure}
\usepackage{amsmath}

\newcommand{\be}{\begin{eqnarray}}
\newcommand{\ee}{\end{eqnarray}}

\def\Io{{\mathbb I}}

\def\sn{\,{\rm sn}\,}
\def\cn{\,{\rm cn}\,}
\def\dn{\,{\rm dn}\,}
\def\sech{\,{\rm sech}\,}

\title{Analytic solutions for a three-level system in a time-dependent field}

\author{Jan Naudts and Winny O'Kelly de Galway\\
\small Departement Fysica, Universiteit Antwerpen,\\
\small Groenenborgerlaan 171, 2020 Antwerpen, Belgium
}

\begin{document}
\maketitle

\begin{abstract}
This paper generalizes some known solitary solutions of a time-dependent Hamiltonian in two ways: The time-dependent
field can be an elliptic function, and the time evolution is obtained for
a complete set of basis vectors. The latter makes it feasible to consider
arbitrary initial conditions. The former makes it possible to observe a beating
caused by the non-linearity of the driving field.
\end{abstract}

\section{Introduction}
\begin{figure}[!h!t]
	\centering
	\includegraphics[width=.9\textwidth]{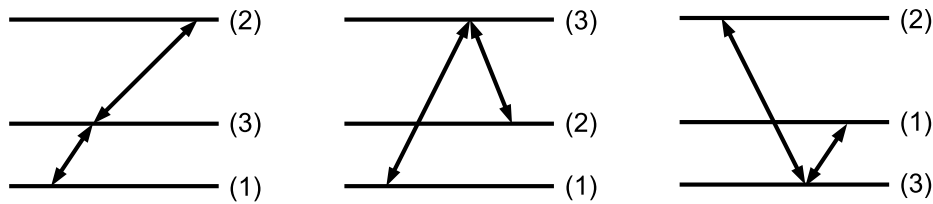}
	\caption{Ladder (or cascade) configuration (left), $\Lambda$ configuration (center),
	and Vee  configuration (right).}
	\label {3level}
\end{figure}

The analytic solutions of Allen and Eberly \cite {AE87}
for the Bloch equations are well-known. Similar results for
spin-one systems or three-level atoms do exist \cite {HFT83} and
are derived in terms of the coherence vector of Hioe and Eberly \cite {HE81}.
We consider a three-level system with time-dependent external fields which enable
transitions between two pairs of levels, between (1) and (3), respectively
between (2) and (3). See the Figure \ref {3level}.
This kind of system has applications in different domains of physics.
Analytic expressions for the time evolution of the density matrix
are for instance very helpful for understanding many of the phenomena observed in
light scattering experiments --- see for instance \cite {FIM05}. In the context of quantum computers the accurate
manipulation of the state of a quantum system --- in this case a qutrit --- is
important.

In the present work the solitary solutions of \cite {HFT83} are generalized
in more than one way. The external fields are modulated with Jacobi's elliptic functions.
By varying the elliptic modulus $k$ these functions make the bridge between
periodic functions ($\cos(\omega t)$ and $\sin(\omega t)$) and
single pulses described by $\sech(\omega t)$ and/or $\tanh(\omega t)$.
In addition, a full set of solutions is presented
instead of just one solution. This makes it possible to take arbitrary initial conditions
at time $t=0$.

The next Section presents the time-dependent Hamiltonian and the special solutions.
In Section 3, a specific setting is chosen. Section 4 discusses the results. The Appendix A
contains the explicit expressions which are used for the generators of SU(3).
The Appendix B explains the method by which the special solutions were obtained
from known solutions (see the appendix of \cite {CKLN00}) of the non-linear von Neumann equation
\be
i\hbar \frac {{\rm d}\,}{{\rm d}t}\rho_t=[H_0,\rho^2_t].
\ee

Next, part of the theoretical framework of \cite {KN09}
was used to obtain a set of linearly independent solutions.
Finally, the results were transferred to a more general setting.

%%%%%%%%%%%%%%%%%%%%%%%%%%%%%%%%%%%%%%%%%%%%%%%%%%%%%%%%%%%%%%%%%%%%%%
\section{Special solutions}

Consider a Hamiltonian of the form
\be
H=H_0+a\cn(\omega t,k)[S_4]_t
+x\dn(\omega t,k)[S_7]_t.
\ee
where  $S_1,S_2,\cdots,S_8$ are the generators of SU(3)
and equal half the Gell-Mann matrices --- see the Appendix A ---
and where
\be
[S_j]_t\equiv e^{-(it/\hbar) H_0}S_je^{(it/\hbar) H_0}
\ee
are the generators written in the interaction picture.
This kind of Hamiltonian is considered in quantum optics when
studying three level systems driven by laser light,
neglecting damping effects --- see for instance \cite {FIM05,CTDRG92}.

The functions $\sn,\cn,\dn$ are Jacobi's elliptic functions.
In the limit $k=0$ the function $\sn(\omega t,k)$ converges to $\sin(\omega t)$,
$\cn(\omega t,k)$ converges to $\cos(\omega t)$, and
$\dn(\omega t,k)$ converges to $1$. In the limit $k=1$ the function
$\sn(\omega t,k)$ converges to $\tanh(\omega t)$ and $\cn(\omega t,k)$
and $\dn(\omega t,k)$ both converge to $\sech(\omega t)$.
In what follows, we drop the arguments $(\omega t,k)$ of the Jacobi functions
when this does not lead to ambiguities.

Let us assume that the parameters of the Hamiltonian satisfy
\be
4k^2(\hbar\omega)^2=a^2+k^2x^2.
\label {maincond}
\ee
Then three orthonormal solutions $\psi_0,\psi_+,\psi_-$ of the Schr\"odinger equation
$i\hbar \dot\psi=H\psi$ are given by
\be
\psi_0(t)
&=&\frac 1Te^{-(it/\hbar) H_0}
\left(\begin{array}{c}
i a\dn\\k^2x\cn\\B\sn\\
\end{array}\right),
\label {tds1}\\
\psi_\pm(t)
&=&\frac {e^{\mp i\phi(t)}}{R(t)}e^{-(it/\hbar) H_0}
\left(\begin{array}{c}
k^2xT\cn\pm iaB\sn\dn\\\pm k^2xB\sn\cn+iaT\dn\\\mp k^4x^2\cn^2\mp a^2\dn^2
\end{array}\right)
\label {tds23}
\ee
with
$B=2k^2\hbar\omega$
and $T=\sqrt{a^2+k^4x^2}$.
Note that by assumption one has $T^2=B^2+a^2(1-k^2)$.
The functions $R(t)$ and $\phi(t)$ are given by
\be
R(t)&=&\sqrt 2\,T\sqrt{T^2-B^2\sn^2}=\sqrt 2\,T\sqrt{a^2(1-k^2)+B^2\cn^2}\\
\hbar\phi(t)&=&\frac {ax}2T\int_0^t{\rm d}s\,\frac {1-k^2}{a^2(1-k^2)+B^2\cn^2(\omega s,k)}.
\ee
One verifies the above statements by explicit calculation.

%%%%%%%%%%%%%%%%%%%%%%%%%%%%%%%%%%%%%%%%%%%%%%%%%%%%%%%%%%%%%%%%%%%%%%
\section{Exciting the ground state}

Let us consider a wave function $\psi(t)$ which at $t=0$
satisfies $\psi(0)=(1,0,0)^{\rm T}$. It can be decomposed into
the basis of special solutions as
\be
\psi(0)
&=&-i\frac aT\psi_0(0)+\frac {k^2 x }{\sqrt 2\,T}\left(\psi_+(0)+\psi_-(0)\right).
\ee
The time-dependent solution is then
\be
\psi(t)
&=&-i\frac aT\psi_0(t)+\frac {k^2 x }{\sqrt 2\,T}\left(\psi_+(t)+\psi_-(t)\right)\crcr
&=&e^{-(it/\hbar) H_0}
\bigg[\frac {a}{T^2}\left(\begin{array}{c}
a\dn\\-ik^2x\cn\\-iB\sn\\
\end{array}\right)
+\sqrt 2\,\frac {k^2 x}{R(t)}\cos(\phi(t))\left(\begin{array}{c}
k^2 x\cn\\ia\dn\\0\\
\end{array}\right)\crcr
& &
+i\sqrt 2\,\frac {k^2 x}{TR(t)}\sin(\phi(t))
\left(\begin{array}{c}
-iaB\sn\dn\\-k^2xB\sn\cn\\k^4x^2\cn^2+a^2\dn^2\\
\end{array}\right)
\bigg].
\ee
Clearly, all three independent solutions are needed to obtain the
time evolution for the given initial condition.
It is also clear that the phase factor $e^{\mp i\phi(t)}$ which appears in
(\ref {tds23}) when $k\not=1$, although not so relevant for the special solutions, becomes
highly relevant in the above quantum superposition.

\begin{figure}[!h!t]
	\centering
	\includegraphics[height=8cm]{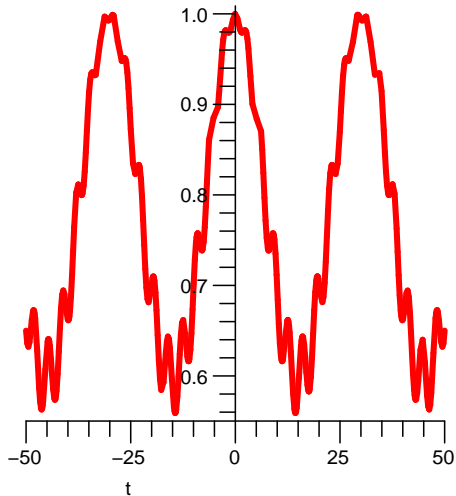}
	\caption{Occupational probability of level (1) as a function of time
for $k=0.25$, $\hbar=\omega=1$, $a=0.3$ and $x=1.6$.}
\label {run3}
\end{figure}
\begin{figure}[!h!t]
	\centering
	\includegraphics[height=7.5cm]{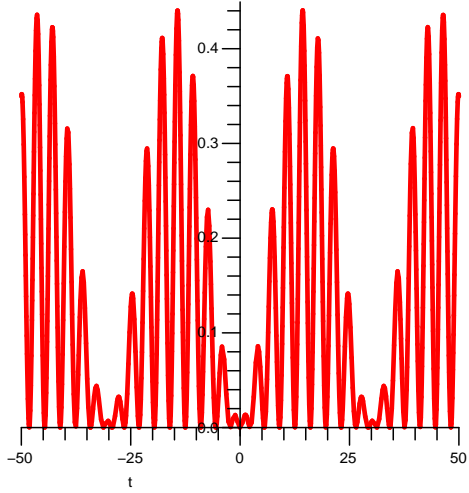}
	\caption{Occupational probability of level (3) as a function of time
for the same parameter values as in the previous Figure.}
\label {run4}
\end{figure}

As expected, the time-dependent interaction populates the
two other states. See the Figures \ref {run3} and \ref {run4}.
Note that level (1) does not go below half occupation.
%This was observed before in ...

%%%%%%%%%%%%%%%%%%%%%%%%%%%%%%%%%%%%%%%%%%%%%%%%%%%%%%%%%%%%%%%%%%%%%%
\section{Discussion}

We obtained solitary solutions for a three-level system with periodic
time-dependent external fields. Two aspects are novel. The external
fields are non-linear in the sense that Jacobi's elliptic functions are used
as deformations of the usual harmonic functions. In addition, a full set of special
solutions is obtained so that arbitrary initial conditions can be considered.

\begin{figure}[!h!t]
	\centering
	\includegraphics[height=5cm]{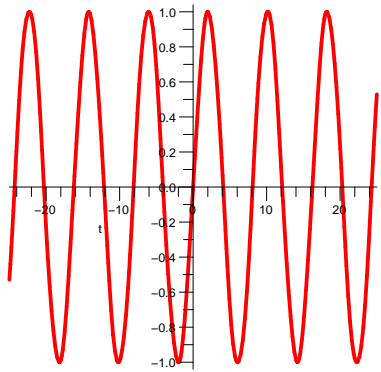}
	\caption{The function $\sin(\phi(t))$ 
for the same parameter values as in the previous Figures.}
\label {phase}
\end{figure}

The additional phase factor $\exp(\mp i\phi(t))$ appearing in the solutions (\ref {tds1}, \ref {tds23})
was first considered in \cite {KN09}. It is not very relevant for the special solutions themselves,
but has effect on their superpositions. The function $\sin(\phi(t))$ is plotted in
the Figure \ref {phase}. Its frequency is slightly lower than the frequency $\omega/2\pi$
of the driving field. As a consequence,
a low frequency beat appears in the case of a superposition of the special solutions.
This is dominantly visible in the Figures \ref {run3}, \ref {run4}.
For the chosen set of parameters the beat period is about 5 times the frequency
of the external field. Note also the frequency doubling by which the exchange
of population occurs between levels (2) and (3).

%%%%%%%%%%%%%%%%%%%%%%%%%%%%%%%%%%%%%%%%%%%%%%%%%%%%%%%%%%%%%%%%%%%%%%
\section*{Acknowledgments}

We acknowledge discussions with Dr.~Maciej Kuna about solving the non-linear von
Neumann equation in the SU(3)-case.

%%%%%%%%%%%%%%%%%%%%%%%%%%%%%
\section*{Appendix A}

The following expressions are used for the generators of SU(3).
\be
S_1&=\frac 12\left(\begin{array}{lcr}
0 &1 &0\\
1 &0 &0\\
0 &0 &0\\
\end{array}\right),
\quad
S_2&=\frac 12\left(\begin{array}{lcr}
0 &-i &0\\
i &0 &0\\
0 &0 &0\\
\end{array}\right),\crcr
S_3&=\frac 12\left(\begin{array}{lcr}
1 &0 &0\\
0 &-1 &0\\
0 &0 &0\\
\end{array}\right),
\quad
S_4&=\frac 12\left(\begin{array}{lcr}
0 &0 &1\\
0 &0 &0\\
1 &0 &0\\
\end{array}\right),\crcr
S_5&=\frac 12\left(\begin{array}{lcr}
0 &0 &-i\\
0 &0 &0\\
i &0 &0\\
\end{array}\right),
\quad
S_6&=\frac 12\left(\begin{array}{lcr}
0 &0 &0\\
0 &0 &1\\
0 &1 &0\\
\end{array}\right),\crcr
S_7&=\frac 12\left(\begin{array}{lcr}
0 &0 &0\\
0 &0 &-i\\
0 &i &0\\
\end{array}\right),
\quad
S_8&=\frac 1{2\sqrt 3}\left(\begin{array}{lcr}
1 &0 &0\\
0 &1 &0\\
0 &0 &-2\\
\end{array}\right).
\ee

%%%%%%%%%%%%%%%%%%%%%%%%%%%%%
\section*{Appendix B}

The solutions (\ref {tds1}, \ref {tds23}) were obtained starting from a known
solution of the non-linear von Neumann equation
\be
i\hbar\dot\rho_t=\frac 32[\{H_0,\rho_t\},\rho_t],
\label {nlvne}
\ee
where $H_0$ is given by
\be
H_0=\frac 23\left(\begin{array}{lcr}
-\mu &0 &0\\
0 &\mu &0\\
0 &0 &\lambda
\end{array}\right).
\label {hnot}
\ee
The three different configurations, vee, ladder, and Lambda, are obtained by taking
$\lambda<-\mu<0$, $|\lambda|<\mu$, respectively $\lambda>\mu>0$.

Let $H(t)$ be defined by $H(t)=\frac 32\{H_0,\rho_t\}$.
Then $\rho(t)$ is a solution of the linear von Neumann equation
with time-dependent Hamiltonian $H(t)$.

A known solution of the non-linear equation (\ref {nlvne}) is of the form \cite {CKLN00,NK06}
\be
\rho(t)&=&
\frac 13\Io
+A\cn(\omega t,k)[S_4]_t
+B\sn(\omega t,k)[S_1]_t
+C\dn(\omega t,k)[S_7]_t\crcr
&=&e^{-(it/\hbar) H_0}\left(\begin{array}{lcr}
\frac 13 &\frac 12B\sn &\frac 12 A\cn\\
\frac 12 B\sn &\frac 13 &-\frac i2 C\dn\\
\frac 12 A\cn &\frac i2 C\dn &\frac 13
\end{array}\right)e^{(it/\hbar) H_0},
\label {specsol}
\ee
The coefficients $A$, $B$, and $C$, are real. They must satisfy the set of conditions
\be
\hbar\omega B
&=&\mu AC\crcr
2 \hbar\omega k^2 C
&=&(\lambda+\mu)AB\crcr
-2 \hbar\omega A
&=&(\lambda-\mu)BC.
\label {appBcond}
\ee
This set of equations can be solved in a straightforward manner when $0<|\lambda|<\mu$.

Next, a unitary matrix $V(t)$, satisfying
\be
\rho_t=V(t)\rho_0V(t)^\dagger.
\label {Vdef}
\ee
is calculated.
The fastest way to find $V(t)$ is by first diagonalizing $\rho_t$.
Note that the eigenvalues of $\rho_t$ do not depend on time. They are given by
\be
\frac 13\quad\mbox{ and }\quad \frac 13\mp \frac 12 T
\ee
with $T=\sqrt{A^2+C^2}$. The result is
\be
V(t)=e^{-(it/\hbar) H_0}G(t)G(0)^\dagger
\ee
with
{\small
\be
& &G(t)=\frac 1{\sqrt {2}TR(t)}\cr
& \times&
\left(\begin{array}{lcr}
i\sqrt 2 CR(t)\dn&-AT\cn-iBC\sn\dn&AT\cn-iBC\sn\dn \\
-\sqrt 2 AR(t)\cn &AB\sn\cn+iCT\dn&AB\sn\cn-iCT\dn \\
\sqrt 2 BR(t)\sn &R(t)^2 &R(t)^2\\
\end{array}\right)
\label {Gdef}
\ee
}
and
\be
R(t)=\sqrt{A^2\cn^2+C^2\dn^2}=\sqrt{T^2-B^2\sn^2}.
\ee

However, $V(t)$ does not necessarily describe the unitary
time evolution $U(t)$. But the latter can be related to $V(t)$
by the method of \cite {KN09}. The knowledge of $U(t)$ implies the
time evolution of wavefunctions $\psi(t)$ for arbitrary initial
conditions $\psi(t)=U(t)\psi(0)$. It turns out that the special solutions
(\ref {tds1}, \ref {tds23}) are the columns of the matrix $G(t)$,
taken in the interaction picture. Two of the three solutions
are multiplied with the time-dependent phase factor $\exp(\mp i\phi(t))$.
Finally, note that the conditions (\ref {appBcond}) are needed during the above
derivation but are not required for (\ref {tds1}, \ref {tds23}) to hold.
They rather are replaced by the single condition (\ref {maincond}).

%%%%%%%%%%%%%%%%%%%%%%%%%%%%%
\section*{}

\end{document}